\documentclass[10pt,twocolumn,letterpaper]{article}

\usepackage{wacv}
\usepackage{times}
\usepackage{epsfig}
\usepackage{graphicx}
\usepackage{amsmath}
\usepackage{amssymb}
\usepackage{booktabs}
\usepackage{float}
\usepackage{placeins}
\usepackage{dblfloatfix}
\usepackage{subcaption}
\usepackage{cuted}

\def\wacvPaperID{438} 

\wacvalgorithmstrack   

\wacvfinalcopy 

\ifwacvfinal
\usepackage[breaklinks=true,bookmarks=false]{hyperref}
\else
\usepackage[pagebackref=true,breaklinks=true,colorlinks,bookmarks=false]{hyperref}
\fi

\begin{document}

\title{On the Importance of Denoising when Learning to Compress Images}



\author{Benoit Brummer\\
intoPIX, University of Louvain\\
Mont-Saint-Guibert, Belgium\\
{\tt\small benoit.brummer@uclouvain.be}
\and
Christophe De Vleeschouwer\\
University of Louvain\\
Louvain-la-Neuve, Belgium\\
{\tt\small christophe.devleeschouwer@uclouvain.be}
}

\maketitle
\thispagestyle{empty}

\begin{abstract}
Image noise is ubiquitous in photography. However, image noise is not compressible nor desirable, thus attempting to convey the noise in compressed image bitstreams yields sub-par results in both rate and distortion. We propose to explicitly learn the image denoising task when training a codec. Therefore, we leverage the Natural Image Noise Dataset, which offers a wide variety of scenes captured with various ISO numbers, leading to different noise levels, including insignificant ones. Given this training set, we supervise the codec with noisy-clean image pairs, and show that a single model trained based on a mixture of images with variable noise levels appears to yield best-in-class results with both noisy and clean images, achieving better rate-distortion than a compression-only model or even than a pair of denoising-then-compression models with almost one order of magnitude fewer GMac operations.
\end{abstract}

\section{Introduction}
\label{sec:intro}

Image sensors capture noise along with useful image information. This noise increases with the camera's ISO sensitivity setting, but noise is virtually always present to some extent and it is both incompressible and undesirable. Lossy image compressors inherently perform some image denoising because removing random noise is often the most effective way to reduce entropy in a signal, but without proper training (or algorithm design) the resulting image size is still inflated and the results look sub-par, as shown in \autoref{fig:noiseeverywhere} (and also attested by Figure S1 in Supplementary Material, and by our experiments in \autoref{fig:mainone}). This increase in bitrate is readily observable in both conventional and learned codecs. A learned lossy image compression scheme is trained by forwarding the image through an autoencoder (AE) and backpropagating from the loss, whose components are the bitrate and the distortion \cite{balle2017}. The bitrate is computed from an optimized, i.e. trained, cumulative distribution function, and the distortion quantifies the difference between the output of the autoencoder and the input image, typically by computing the mean square error (MSE).

\begin{figure*}
  \begin{center}
    \includegraphics[width=\linewidth]{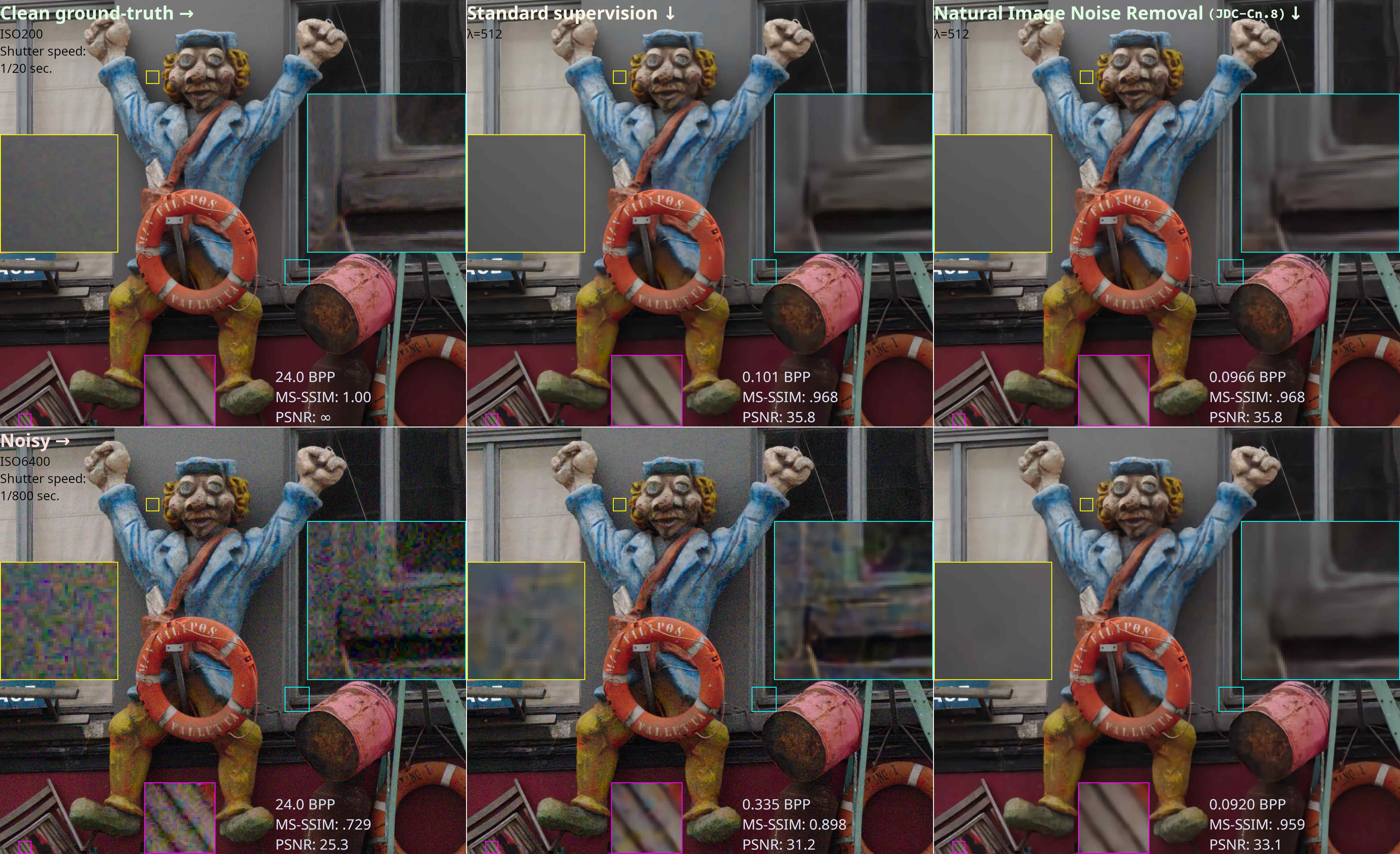}
    \caption{Visualization of a clean (top) and noisy (bottom) test image from NIND \cite{nind}. From left to right: (i) ground-truth/noisy input, (ii) compression autoencoder trained with standard supervision \cite{manypriors}, relying on the adoption of a perception-based loss function to mitigate the impact of noise, (iii) our proposed joint denoising and compression model trained with Natural Image Noise Removal supervision using both clean and low noise images (``JDC-Cn.8'').}
    \label{fig:noiseeverywhere}
  \end{center}
\end{figure*}

Any image compression scheme can attain better rate-distortion by having the noise removed first.
An image denoiser can typically be trained to reconstruct clean images from noisy inputs, using a dataset of paired images where static scenes are captured using a progressively faster shutter speed \cite{nind}.

In this work, we consider joint compression and denoising. Adding a denoising functionality essentially comes down to feeding the network with a potentially noisy image and comparing its output with a clean image that may have a better quality than what was initially input to the network. The goal is to generate an image that is of higher quality than the one used as input, while decreasing the necessary bitrate close to that of a clean image. Meanwhile, there is no added complexity because the inference process and the network architecture remain unchanged. The network is trained with both noisy images and some clean images as input such that it removes noise while retaining the ability to compress clean input images efficiently.

Our experiments analyzes the impact of image noise on the rate-distortion of different standard and learned compression methods, the benefit of performing denoising prior to compression, and denoising while compressing. Our original supervision strategy, introduced to promote the reconstruction of clean images when the learned codec is fed with noisy ones, appears to be effective. 
The resulting joint denoising and compression models perform properly on clean images as well as noisy ones, effectively replacing a standard compression models for general-purpose image compression and substantially improving the rate-distortion on noisy images. As illustrated in the second line of \autoref{fig:noiseeverywhere} (comparison between second and third columns), it is shown to significantly improve rate-distortion performance (using non-noisy images as ground-truth) compared to relying on the implicit noise removal induced by the conventional adoption of a perception-based loss function during training. It also reaches slightly better rate-distortion than a computationally heavy two-step procedure, involving one AE for denoising, followed by another AE-based network for compression.

This paper is organized as follows: \hyperref[sec:background]{Section 2} summarizes the work on which this paper builds. The main concepts behind our joint denoising and compression supervision are introduced in \hyperref[sec:dc]{Section 3}. The implementation details are given in \hyperref[ssec:methods]{Section 4} followed by the results, and \hyperref[sec:conclusion]{Section 5} summarizes the impact of the present work.

\section{Background}
\label{sec:background}


\textbf{Learned Lossy Image Compression} is typically based on the seminal work of Johannes Ballé et al. \cite{balle2017}; a convolutional autoencoder \cite{autoencoder} with generalized divisive normalization (GDN) \cite{gdn}, and an entropy model which is jointly optimized to capture the latent distribution. This model has been extended with a 
parametrized hyperprior \cite{balle2018} or with a competition between multiple priors \cite{manypriors}, which allows for manipulating an image-dependent latent distribution. Our experiments build onto the initial architecture from \cite{balle2017} completed with multiple sets of latent distributions learned in \cite{manypriors}.

\textbf{Image Noise} occurs as the sensitivity of an image sensor is increased to make up for non-ideal lighting conditions. When insufficient light is provided or the dynamic range is too wide, the ISO and/or shutter speed settings are increased accordingly. Pixels take on random, less accurate values, and fewer detail is visible as a result. Different image restoration techniques have been developed to tackle image denoising, including Wavelet \cite{wavdenoise,wavdc} and non-local means based methods \cite{nlm}, BM3D \cite{bm3d}, and recent deep-learning based methods \cite{nind,learningtoseeinthedark,cnndenoise1,cnndenoise2}. Image noise does not reflect the physical components of the observed scene. It is a consequence of the imperfect acquisition process, and thereby should be ignored when possible. Hence, \textbf{targeting the reconstruction of a denoised image is the proper way to proceed to get a faithful representation of reality} (even if it implies to not perfectly render the captured signal).

The \textbf{Natural Image Noise Dataset} (NIND) \cite{nind} and Smartphone Image Denoising Dataset (SIDD) \cite{sidd} provide sets of clean--noisy image pairs which are appropriate to train a denoising neural network. NIND is made of multiple pictures of many static scenes captured on a tripod to ensure spatial consistency; the clean ground-truth images are taken in ideal conditions with a camera's base ISO sensitivity to capture as much light as is necessary and to obtain the best possible quality, and matching versions of the same scene are captured with a variety of increasing shutter speed and ISO settings, which result in increased noise and lower image quality. These noisy images are typically fed as the input of the denoising neural network while training it to reconstruct the scene as if it were taken in ideal conditions.

\textbf{Denoising and Compression} have been studied as a combined problem in the wavelet domain \cite{wavdc,wavdc2}, 
and more recently the idea of learning denoising and compression jointly with neural networks was approached by Testolina et al. \cite{testolina} in the context of the JPEG AI codec. The decoder in \cite{testolina} is extended such that it consists of twice as many layers, and Poissonian-Gaussian noise is applied to the input training data. This approach delegates the denoising task to the decoder, in line with the JPEG AI requirements of using a universal encoder and specialized decoders. However, as shown in our experimental section, this architecture results in no appreciable bitrate reduction because it is only the encoder that can ensure that incompressible noise does not reach the bitstream. Moreover, training a denoiser with synthetic noise tends to produce a poor model on real data \cite{nind,dnd}. Testolina et al. introduce a promising joint denoising and compression (JDC) scheme but the resulting rate-distortion of their model falls short of that obtained using our proposed supervision based on pairwise naturally noisy / clean images.

\section{Jointly Learned Denoising and Compression}
\label{sec:dc}

An effective denoising network can be trained to reconstruct clean images given noisy images as input, using a dataset of paired noisy--clean images such as NIND \cite{nind} and SIDD \cite{sidd}. We propose to adopt a similar principle to train an autoencoder originally designed for image compression.

\begin{figure*}
  \begin{center}
    \includegraphics[width=\linewidth]{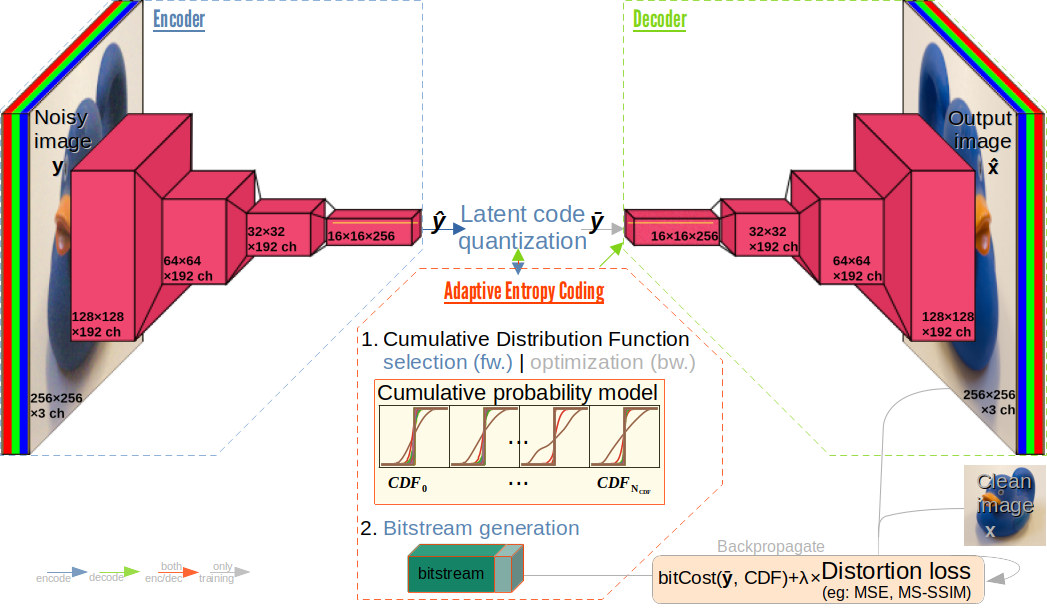}
    \caption{Denoising and compression joint training: the distortion loss is computed between the reconstructed image $\hat{x}$ and a clean image $x$. The input image $y$ may be noisy. The network \cite{manypriors} is made of four (transposed) convolutions with stride of 2 and kernel size of 5, each followed by a GDN activation \cite{gdn} except for the last (transposed) convolution. (Best viewed in color.)}
    \label{fig:compdenoise_schema}
  \end{center}
\end{figure*}

A joint denoising and compression autoencoder \cite{manypriors} is trained to generate a clean image from either a matching noisy image in a paired dataset or from the same clean image. The aim is to obtain a decoded image whose quality is potentially higher than that of the input image, while saving the space that would otherwise be wasted in encoding noise. Different methods are proposed to train such a joint denoising and compression model using Natural Image Noise Removal (NINR) supervision. They are described in Section \ref{ssec:practicalimplementations} and \autoref{fig:compdenoise_schema} illustrates the general training process.

\subsection{Our Proposed NIN Supervision Strategies}
\label{ssec:practicalimplementations}


Four different strategies are envisioned and compared to implement our novel Natural Image Noise Removal (NINR) supervision paradigm. They are listed in \autoref{tab:methodsacro} and described below.

\textbf{Noisy Pairs (JDC-N)} The simplest joint denoising and compression implementation consists of training with all noisy--clean image pairs available in the dataset(s).

\textbf{Clean and Noisy Pairs (JDC-CN)} This method considers some clean--clean image pairs, in addition to the noisy--clean image pairs, to ensure that the network's performance does not degrade when the input images contain no unwanted noise. The dataset of clean images can be selected as a set of images which has been assessed and promoted by human reviewers, such as the Wikimedia Commons Featured Pictures \cite{fp}, then further refined by eliminating images whose metadata indicates a high ISO value in order to ensure the absence of noise.

\textbf{Clean and Low-noise Pairs (JDC-Cn)} To specialize the model to the most frequent input image noise levels, we have also considered placing a threshold on the training data noise level. Our experiments reveal that it is beneficial to filter out the most noisy input training images because the overall rate-distortion degrades when the network is trained to perform more extreme denoising. Such extreme denoising would require added complexity on the encoder and, although possible, extreme denoising is outside the scope of a combined denoiser whose aim is to improve rate-distortion by removing noise that is inherently present in most photographs rather than learning to see in the dark, as proposed in \cite{learningtoseeinthedark}. The paired image noise dataset is analyzed prior to training such that the multi-scale structural similarity (MS-SSIM) \cite{msssim} score between each noisy crop and its clean ground-truth is stored in a file, and the training dataset can be initialized such that all training crops exceed a set quality threshold.  The effect of different noise thresholds is analyzed in the \hyperref[sssec:ablation]{ablation study}.

\textbf{Building Pairs from a Universal Denoiser (JDC-UD)} A fourth training method consists of running a pre-trained blind denoising model \cite{unet,nind} on all the training data to generate the ground-truth images, and computing the training loss between the input images and the denoised images. This method effectively performs knowledge distillation \cite{knowledgedistillation} from a powerful universal denoising network to the joint denoising and compression network. All input images are considered noisy and the training dataset is virtually limitless because the ground-truth images are generated (in advance), thus entire image datasets are used without filtering.

%
%
%
%
%
%
%
%
%
%
%

\begin{table*}[]
\centering
\caption{Data pairs considered in this paper to train a joint denoising and compression model. JDC-Cn is also referred to with its training noise threshold (e.g. JDC-Cn.8 is trained with $\text{MS-SSIM} \ge 0.8$, see the text for details).}
\label{tab:methodsacro}
\begin{tabular}{|l|l|l|}
\hline
\textbf{Method} & \textbf{Training input}                     & \textbf{Expected output}         \\ \hline
JDC-N           & Noisy image from clean--noisy paired dataset \cite{nind} & Clean ground-truth               \\ \hline
JDC-CN &
  \begin{tabular}[c]{@{}l@{}}Noisy image from clean--noisy paired dataset,\\ clean image from high quality dataset \cite{fp}\end{tabular} &
  \begin{tabular}[c]{@{}l@{}}Clean ground-truth,\\ clean input\end{tabular} \\ \hline
JDC-Cn &
  \begin{tabular}[c]{@{}l@{}}Weakly noisy image from clean--noisy paired dataset,\\ clean image from a high quality dataset\end{tabular} &
  \begin{tabular}[c]{@{}l@{}}Clean ground-truth,\\ clean input\end{tabular} \\ \hline
JDC-UD          & Arbitrary input image \cite{fp,nind}                            & Provided by a universal denoiser \\ \hline
Testolina \cite{testolina}          & Clean image from high quality dataset + artificial noise                             & Clean input \\ \hline
\end{tabular}%
\end{table*}

\section{Experiments}
\label{sec:exp}


\subsection{Practical Implementation Details}
\label{ssec:methods}

These experiments are based on the PyTorch implementation of the autoencoder base codec introduced in \cite{manypriors}. Source code is provided as Supplementary Material and available on \url{https://github.com/trougnouf/compression}. The training loss of the compression autoencoder is computed as $\text{Loss}=\text{bitrate}(\hat{x})+\lambda \times \text{MSE}(\hat{x}, x)$, where $\hat{x}$ is the decoded image and $x$ is the clean ground-truth which, as explained in Section 3, may differ from the input image, and $\lambda$ balances the rate/distortion trade-off of the model. The combined denoising and compression autoencoder \cite{manypriors} is trained with batches of four noisy images from NIND \cite{nind} and one clean image from the Wikimedia Commons Featured Pictures \cite{fp} dataset whose ISO value does not exceed 200, with a crop size of 256 as is typically done in the learned compression literature \cite{manypriors,balle2017,balle2018}. The pre-trained ``\textbf{universal denoiser}'' used to train the JDC-UD model is the U-Net-like blind denoising model published with NIND, which was updated such that its training includes clean--clean image pairs. The CLIC professional test set \cite{clictest} is used to assess the models on other clean images.


\begin{figure*}[b]
\centering
\begin{subfigure}{.5\textwidth}
  \centering
  \includegraphics[width=.9\linewidth]{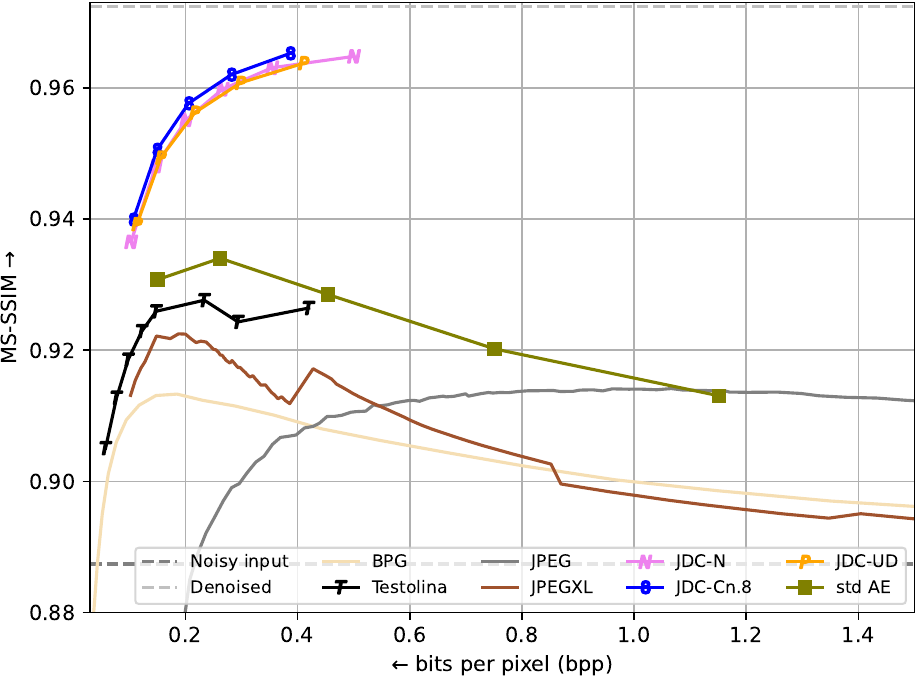}
  \caption{Compression when taking the original (and thus noisy) image \\as input.}
  \label{fig:mainone}
\end{subfigure}%
\begin{subfigure}{.5\textwidth}
  \centering
  \includegraphics[width=.9\linewidth]{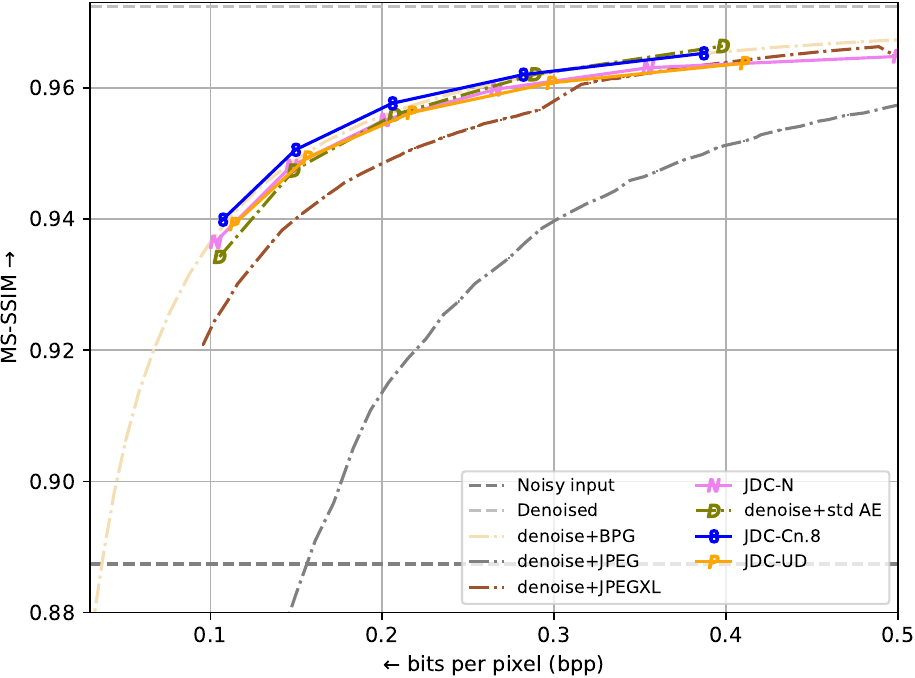}
  \caption{Applying CNN-based denoising before compression \\(except for our JDC which  inputs the originally captured image).}
  \label{fig:maintwo}
\end{subfigure}
\caption{Lossy compression of noisy ($\text{MS-SSIM}\in [0.7, 1.0)$) test images from NIND \cite{nind} with respect to their matching clean ground-truth. \\\textbf{(a)} Original (noisy) images are provided as input. Standard methods (JPEG, JPEG XL, BPG, and a standard compression autoencoder \cite{manypriors}) perform some implicit denoising at low bitrates, but image quality degrades as the bitrate increases since the noisy signal is reconstructed. \\\textbf{(b)} A universal denoiser is applied before presenting the image to the standard method. This greatly improves rate-distortion but adds of an order of magnitude of complexity. Our joint denoising and compression (JDC) autoencoder with Natural Image Noise Removal supervision allows noise-free reconstruction without prior denoising.}
\label{fig:mainres}
\end{figure*}

The JDC model defined in Testolina et al. (2021) \cite{testolina} is trained entirely as described with Poissonian-Gaussian artificial noise \cite{artnoise} (with noise parameters $a=0.2^2, b=0.04^2$), the encoder from Balle et al. (2018) \cite{balle2018}, and their proposed decoder which has twice as many layers as the one recommended by Ballé et al. An additional JDC-Cn model is trained with the larger decoder proposed by Testolina et al. in order to assess their proposed network architecture separately from the training method.

Most models are trained for six million steps with $\lambda=4096$ to yield the highest bitrate, then the $\lambda$ value is halved and training continues for three million steps for each $\lambda$ value all the way down to $\lambda=256$, like in \cite{manypriors}. Both the JDC-Cn.8-Tdec method that is matched with the decoder defined by Testolina et al. and the JDC-N model trained with only noisy--clean image pairs have had an additional model trained with $\lambda=8192$ in order to reach the other methods' highest bitrate. Likewise, the whole method defined by Testolina et al. \cite{testolina} was trained with up to $\lambda=16384$.

The standard codecs comparisons are made by encoding images using GraphicsMagick 1.3.37 (JPEG), the JPEG XL encoder v0.6.1, and the BPG Image Encoder version 0.9.8. The ``\textbf{standard autoencoder}'' is that defined \cite{manypriors}.

\subsection{Results}
\label{ssec:results}

\begin{figure*}[b]
\begin{subfigure}{.5\textwidth}
  \includegraphics[width=.9\linewidth]{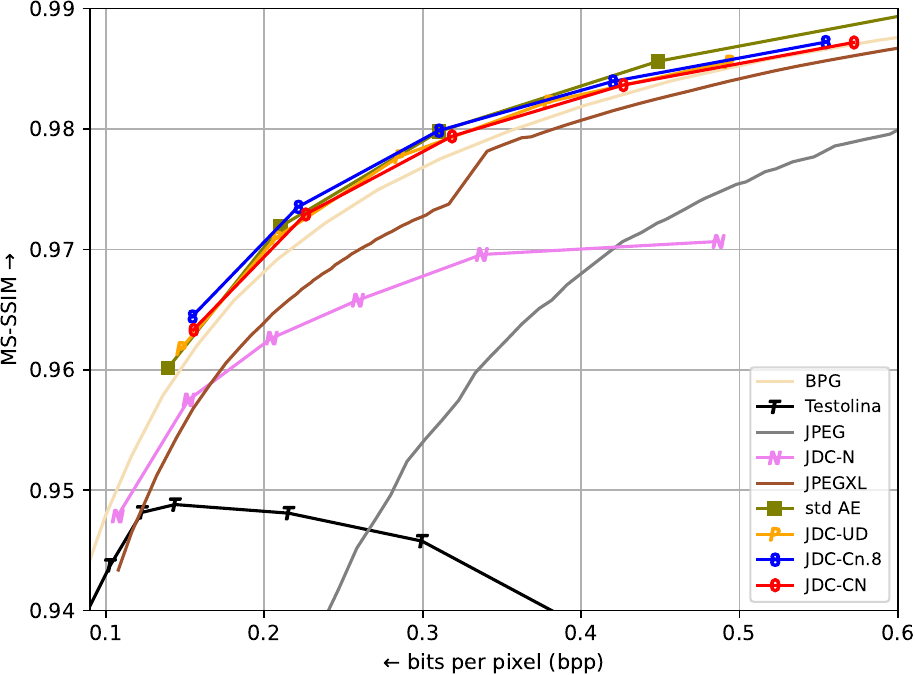}
  \caption{High quality (unpaired) test images}
  \label{fig:clicpro}
\end{subfigure}%
\begin{subfigure}{.5\textwidth}
  \includegraphics[width=.9\linewidth]{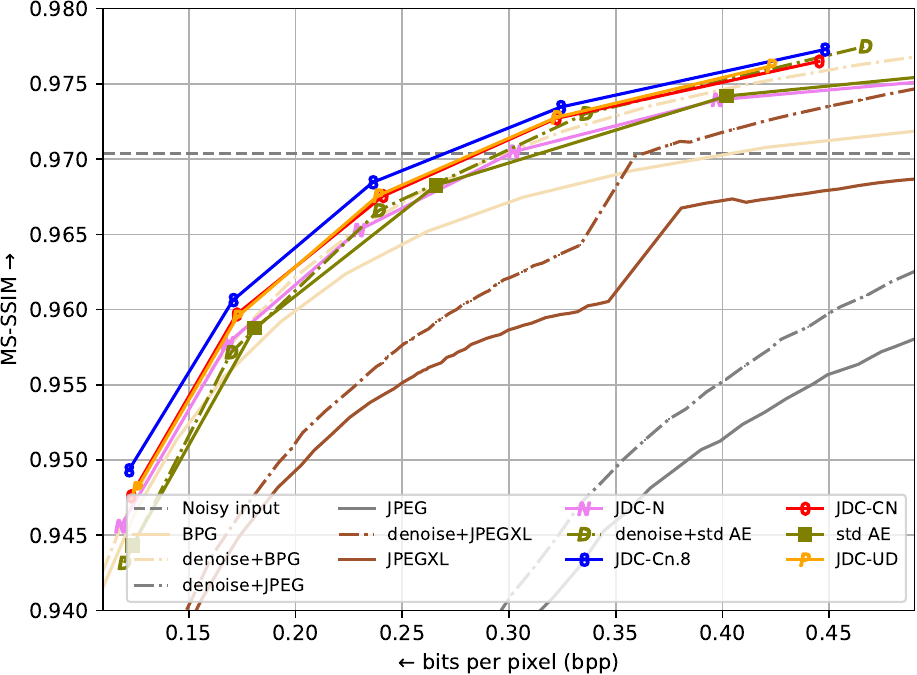}
  \caption{Test images with ($\text{MS-SSIM}\in [0.95, 1.0)$, $\text{MS-SSIM}_\mu=0.97$)}
  \label{fig:hqgraph}
\end{subfigure}
\caption{MS-SSIM rate-distortion curve of different compression methods on (a) high quality images from the CLIC pro test set \cite{clictest} and (b) nearly noiseless test images from NIND \cite{nind}.}
\label{fig:lownoisecomp}
\end{figure*}

\begin{figure*}
\begin{subfigure}{.5\textwidth}
  \includegraphics[width=.9\linewidth]{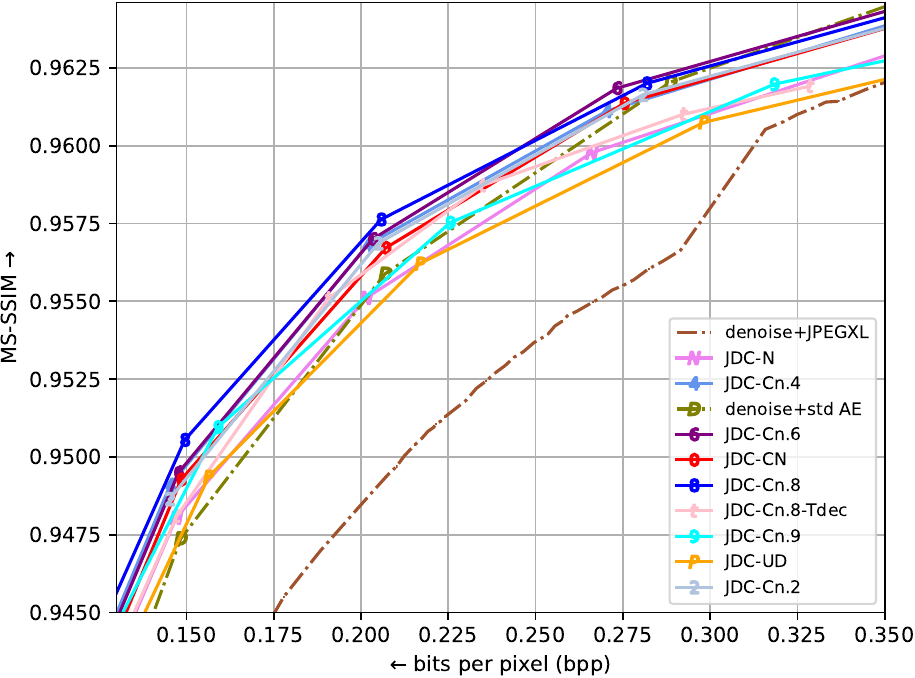}
  \caption{Test images with $\text{MS-SSIM}\in [0.7, 1.0)$, $\text{MS-SSIM}_\mu=0.89$}
  \label{fig:lownoise7}
\end{subfigure}%
\begin{subfigure}{.5\textwidth}
  \includegraphics[width=.9\linewidth]{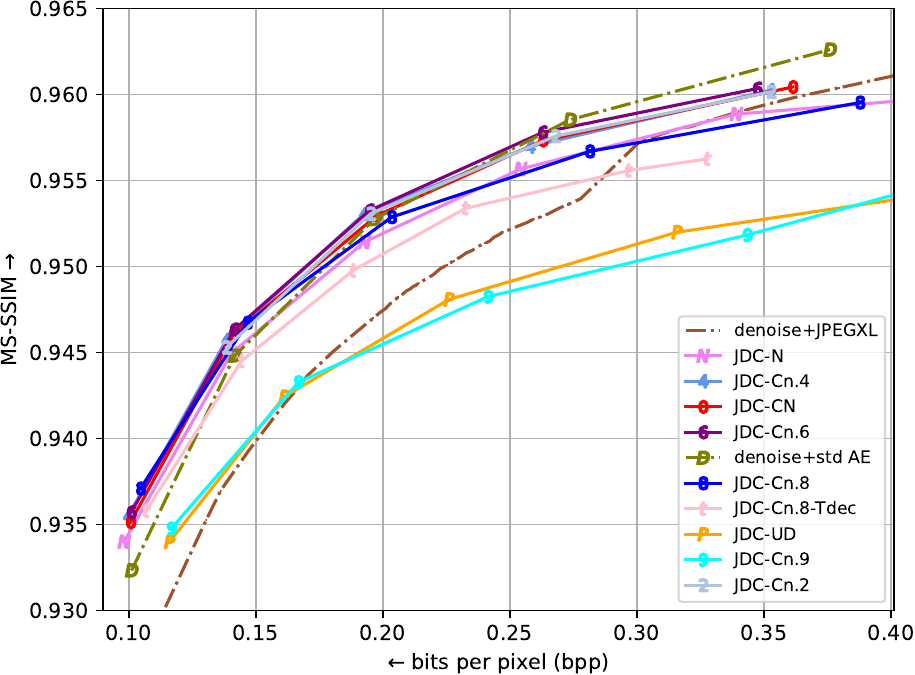}
  \caption{Test images with $\text{MS-SSIM}\in [0.5, 1.0)$, $\text{MS-SSIM}_\mu=0.84$}
  \label{fig:lownoise5}
\end{subfigure}
\caption{Lossy compression of noisy test images from NIND \cite{nind} with different joint denoising and compression methods including JDC-Cn trained with different MS-SSIM thresholds, JDC-CN trained with no such threshold, JDC-N trained with no clean images, JDC-UD trained with knowledge distillation from a universal denoiser, and JDC-Cn.8 trained with the larger decoder defined by Testolina et al. \cite{testolina}. \textbf{(a)} The JDC-Cn methods tend to perform well, especially when the quality threshold is set between 0.6 and 0.8, but the rate-distortion worsens when the training quality threshold increases to 0.9 (i.e. little noise is seen by the model during training). The JDC-UD model yields similarly lower performance, and so does the larger decoder despite using the same training scheme as JDC-Cn.8. \textbf{(b)} The noise level is more extreme than what is likely to occur in photographs. This further shows that compressing with more noise than is ever seen during training (e.g. JDC-Cn.9) yields worse rate-distortion. The model trained without clean image pairs (JDC-N) does not perform as well even when the test images are noisier.}
\label{fig:lownoisecomp}
\end{figure*}

\begin{figure*}
  \centering
    \includegraphics[width=\textwidth]{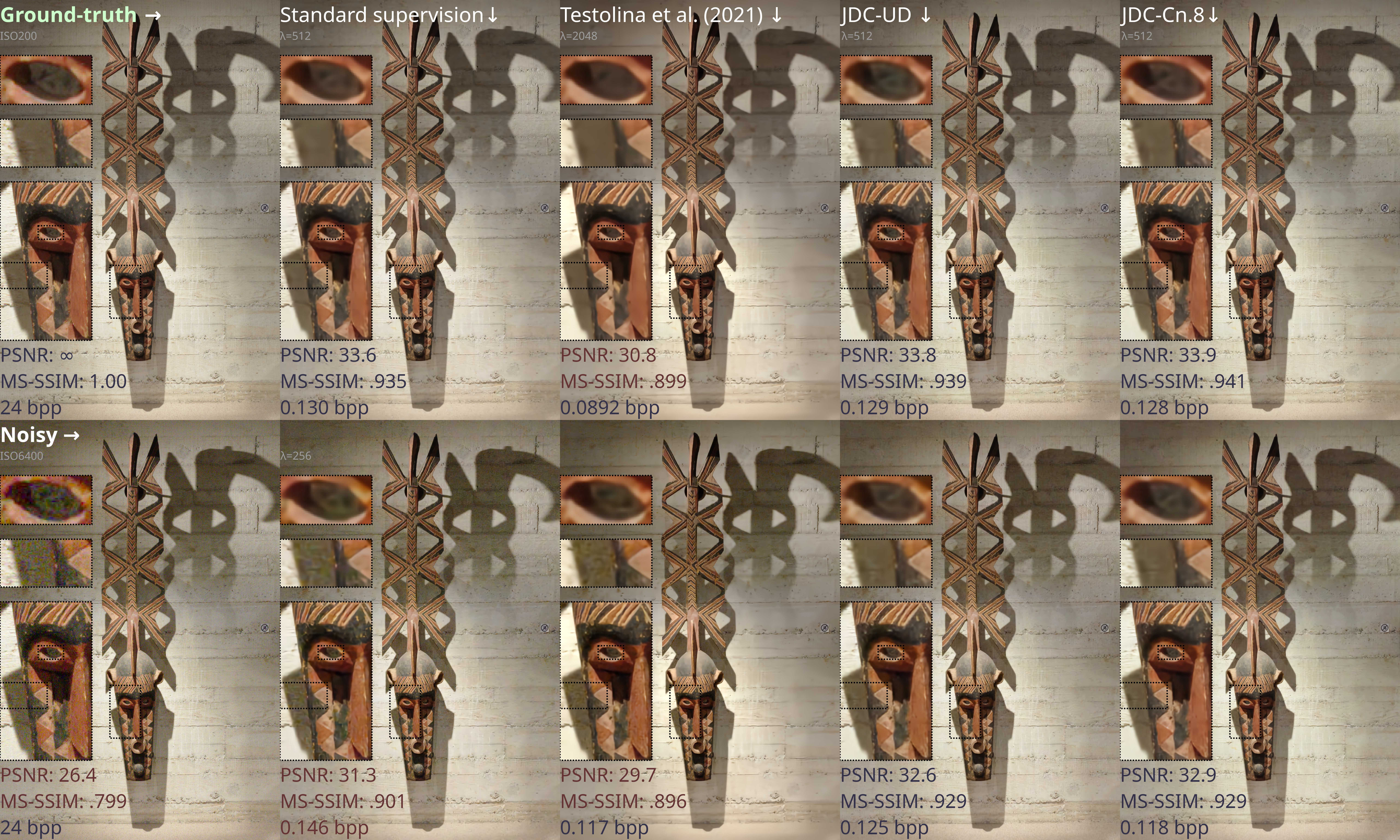}
    \caption{Visualization of a clean (top) and noisy (bottom) test image from NIND \cite{nind} encoded with a target bitrate of 0.13 bpp using different trained compression autoencoders. From left to right: ground-truth/noisy input, standard autoencoder \cite{manypriors}, autoencoder from Testolina et al. (2021) \cite{testolina} (trained on artificial noise; increasing bitrate did not yield quality improvements on the ground-truth), joint model trained with knowledge distillation from a universal denoiser (JDC-UD), and joint model trained with both clean and low noise images (JDC-Cn.8). Images are best visualized after zooming in on a screen.}
    \label{fig:visualcomp}
\end{figure*}

\subsubsection{On the Importance of Denoising.}

The first experiment measures the impact of denoising prior to compression with different compression methods. \autoref{fig:mainone} plots the rate-distortion curves obtained without specific denoising of the input images, for a variety of codecs. We observe that for all codecs compression is an effective denoising method at the lowest bitrate. However, at reasonable bitrates, all conventional codecs (learned or not) tend to reproduce the noise. This is in contrast with our proposed joint denoising and compression paradigm, which continuously increases quality when the bitrate increases.
Denoising before compression might be considered to solve the quality issue when using conventional codecs. As shown in \autoref{fig:maintwo}, this bridges (most of) the quality gap compared to our proposed JDC method, but at the cost of a significantly increased complexity (see Section \ref{sssec:complexity}).

\begin{figure*}[b]
  \centering
    \includegraphics[width=\textwidth]{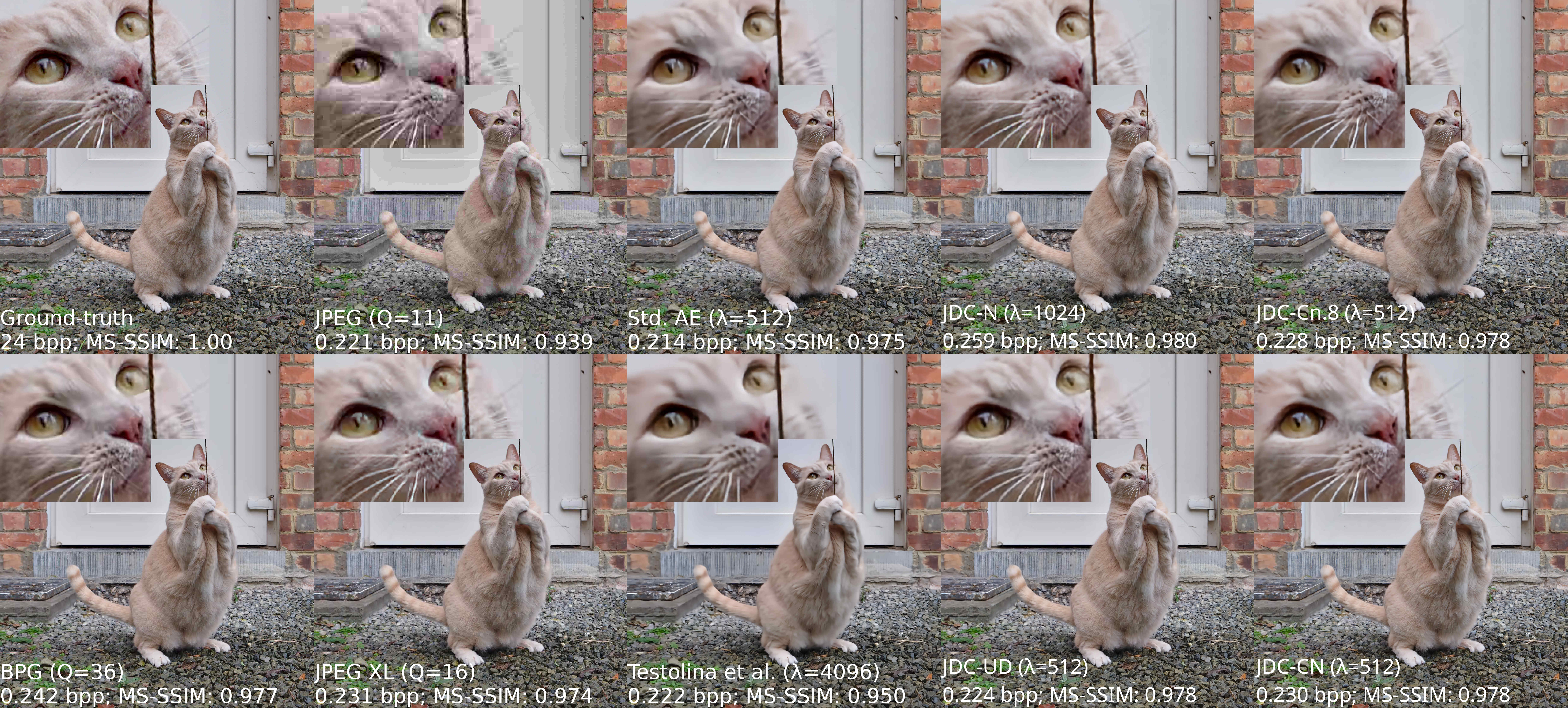}
    \caption{Visualization of a clean test image compressed using different methods with a target bitrate of 0.23 bpp. Standard methods (JPEG, JPEG XL, BPG) tend to produce block artifacts. The method defined by Testolina et al. \cite{testolina} produces oversmoothed results on clean images. Training with only noisy images can produce the same level of quality at the cost of increased bitrate. Other methods---training with only clean images, training with both clean and noisy images, and knowledge distillation from a powerful denoiser---perform well on clean images. The input image has been processed in the darktable software with (among other methods) non-local means \cite{nlm} profiled denoising.}  
    \label{fig:larry512}
\end{figure*}





\subsubsection{Our Joint Denoising and Compression Scheme} \autoref{fig:mainres} also introduces the proposed joint denoising and compression models, JDC-CN (no noise threshold) and JDC-Cn.8 (training noise limited to $\text{MS-SSIM}\ge 0.8$). These models are trained like the dedicated denoising model in that the input training batch is made of four noisy images and one clean image, and the model is tasked with reconstructing the clean ground-truths. This single JDC model generally achieves better rate-distortion than a duo of denoising then compression neural networks (except at high bitrate), while using significantly less computational overhead.

The best results shown are obtained with the ``JDC-Cn.8'' model which is trained with paired images whose input noise is limited to $\text{MS-SSIM} \ge 0.8$ as well as with unpaired clean images to promote generalization. $T$ is the method trained with artificial noise described by Testolina et al. \cite{testolina}, which performs worse than a duo of models as is also shown in their results.

\subsubsection{Computational Complexity}\label{sssec:complexity} Computational cost is measured in terms of billion multiply-accumulate operations (GMac) and runtime on an AMD Threadripper 3960X CPU. The dedicated denoising U-Net performs 812 billion multiply-accumulate operations (GMac) per megapixel (MP) in 65.8 sec./MP, whereas the JDC model's compression encoder performs 92.8 GMAC/MP \cite{manypriors} in 2.9 sec./MP. The dual model approach (denoising then compression) thus performs a total of 904.8 GMac/MP, whereas a single joint denoising and compression model operates with 10.3\% of that computational complexity.

\subsubsection{Handling Clean Images}

JDC-C models are trained with both noisy--clean and clean-clean paired images in order to better generalize and maintain good rate-distortion on clean images. \autoref{fig:clicpro} and \autoref{fig:larry512} show the behavior of different JDC training strategies when compressing clean images. JDC models trained with some clean input images or with the JDC-UD knowledge distillation technique yield a rate-distortion similar to the model trained for compression only, even when no minimum training noise threshold is set, thus incorporating clean images in the training data (JDC-CN) reinstates a good rate-distortion. Limiting the input noise to $\text{MS-SSIM}\ge 0.8$ (JDC-Cn.8) further improves rate-distortion such that it is slightly better with a JDC model than a standard model at low bitrates, and slightly worse at high bitrates due to the perception-distortion tradeoff \cite{perceptiondistortion} where only reconstruction fidelity matters. The JDC-N model trained with only noisy--clean image pairs performs significantly worse on clean images, and the model trained with artificial noise (``$T$'') performs worst.

\autoref{fig:hqgraph} shows a common use-case where the amount of noise is low ($\text{MS-SSIM}\in[0.95, 1)$). Prior denoising still improves rate-distortion, and joint denoising and compression methods yield the most significant rate-distortion benefits. All compression methods benefit from prior or joint denoising even when the level of noise is minor. Traditional compression schemes benefit the most from prior denoising, and joint denoising outperforms prior denoising in learned methods.


\subsubsection{Ablation Study}
\label{sssec:ablation}

The effect of different training noise thresholds is analyzed when compressing noisy images. In Figure \autoref{fig:lownoise7} the test noise is limited to $\text{MS-SSIM}\in[0.7, 1)$ which is qualitatively fairly noisy, as shown in \autoref{fig:noiseeverywhere}. None of the methods perform significantly better or worse than the denoise and compress duo of models. It is worth noting that the three worst JDC training schemes are the knowledge distillation JDC-UD model, the model trained with a quality threshold of $\text{MS-SSIM}\ge 0.9$, and the decoder defined by Testolina et al. which contains twice as many layers. The JDC-Cn models trained with an MS-SSIM threshold of 0.6 and 0.8 yield the best rate-distortion. A visualization of the different denoising and compression methods at low bitrates is shown as \autoref{fig:visualcomp}.


In \autoref{fig:lownoise5}, the testing noise is increased to $\text{MS-SSIM}\in[0.5, 1)$, showing how the models behave under extreme input noise. The results are largely the same; the model trained with $\text{MS-SSIM}\ge 0.9$ struggles even more due to the increased noise and its performance is close to that of the JDC-UD method, the model trained with $\text{MS-SSIM}\ge 0.8$ does not perform as well whereas the model trained with $\text{MS-SSIM}\ge 0.6$ is still competitive, and it remains beneficial to train with clean image pairs as well.


%
%

%
%
%
%
%
%
%
%


\section{Conclusion}
\label{sec:conclusion}

Denoising images improves rate-distortion whenever there is noise present, regardless of the compression method. Denoising can be performed prior to compression using a dedicated denoiser (as is typically done in professional image development workflow) with no adaptation to the compression scheme. A joint model that is trained to perform denoising and compression simultaneously yields further improvements in rate-distortion.
As a result, a joint denoising and compression model performs 8.9 times fewer GMAC operations than a U-Net denoiser followed by a compression encoder. Since the JDC model only differs from standard learned compression models by the adopted supervision strategy, it can be implemented using any of the compression architectures available in the literature (such as \cite{balle2017,balle2018,manypriors}). Our proposed Natural Image Noise Removal supervision strategy thus provides a fundamental and generic contribution that is expected to become popular in future works related to learned compression.

In practice, joint denoising and compression models may be trained using a dataset of noisy--clean image pairs with natural noise, such as NIND \cite{nind} and SIDD \cite{sidd}. Performance is improved by setting a quality threshold on the training images, such as $\text{MS-SSIM} \ge 0.8$ or $\text{MS-SSIM} \ge 0.6$ depending on the maximum expected noise. The rate-distortion curve is preserved on clean images and improved in any case
 by incorporating clean-clean image pairs in the training data.

An alternative method consists of performing knowledge distillation \cite{knowledgedistillation} by using the output of a dedicated denoiser as ground-truth images during training. This has the benefit of allowing a virtually limitless training dataset because a paired dataset is no longer required, but it requires pre-processing of the training images and results in a slightly worse rate-distortion.

\section{Acknowledgements}

This research has been funded by the Walloon Region. Computational resources have been provided by the supercomputing facilities of the Université catholique de Louvain (CISM/UCL) and the Consortium des Équipements de Calcul Intensif en Fédération Wallonie Bruxelles (CÉCI) funded by the Fond de la Recherche Scientifique de Belgique (F.R.S.-FNRS) under convention 2.5020.11 and by the Walloon Region.

%
\section{Annex}

\begin{figure*}
    \centering
    \includegraphics[width=\linewidth]{"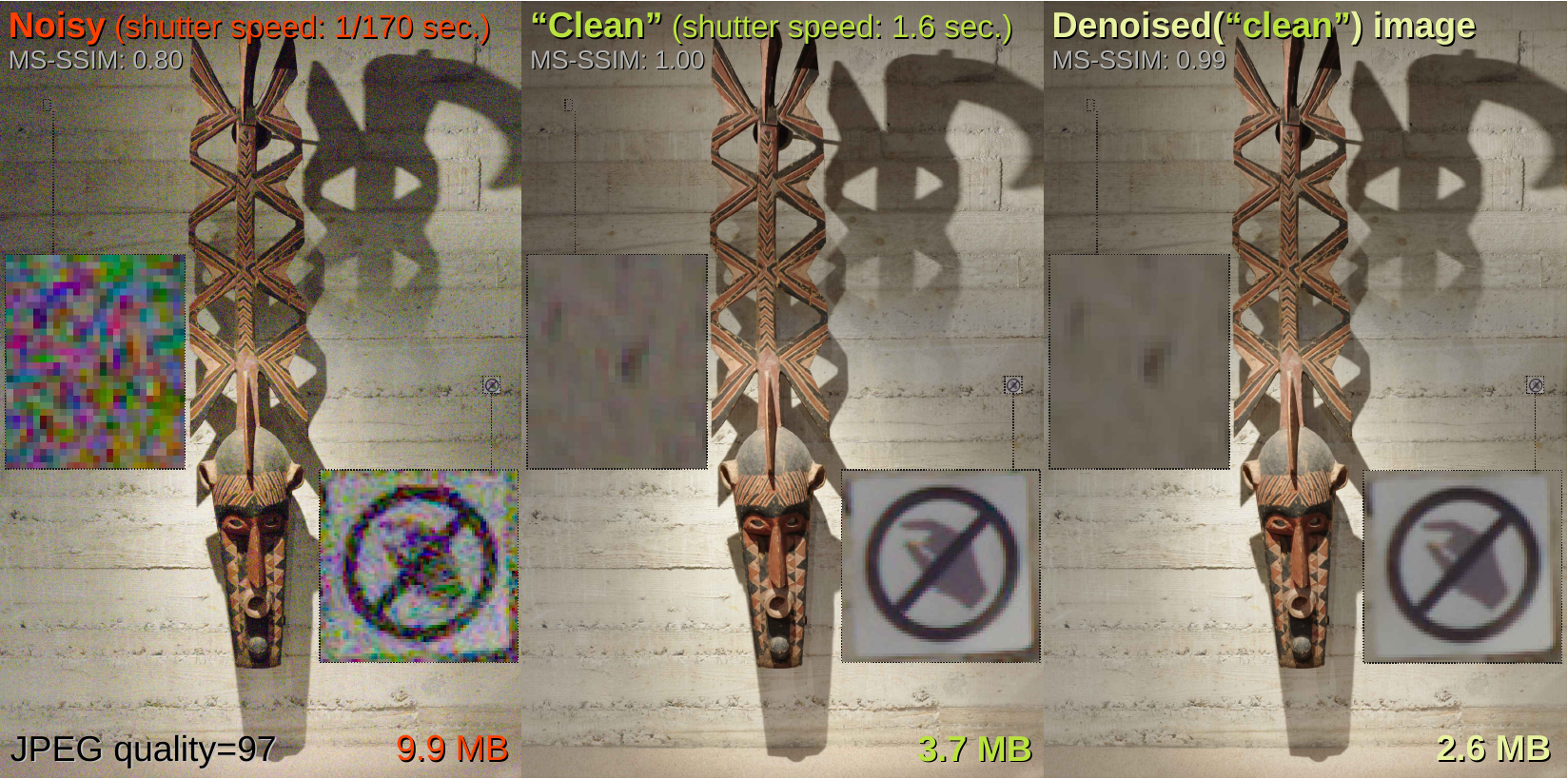"}
    \captionof{figure}{Noise is both incompressible and undesirable, as shown in this clean--noisy image pair from the Natural Image Noise Dataset (NIND) \cite{nind}, where the same scene has been captured with increasingly faster shutter speed. The JPEG compressed (Q=97) image size increases from 3.7 MB for the ground-truth to 9.9 MB when encoding the noisy image. Even the ground-truth image (center) contains some background noise and artifacts such as chromatic aberrations, and JPEG compresses it down to 2.6 MB when it is denoised with a trained denoiser from \cite{nind} prior to compression (left).}
    \label{fig:noiseeverywhere}
\end{figure*}


\begin{figure*}
  \begin{center}
    \includegraphics[width=\linewidth]{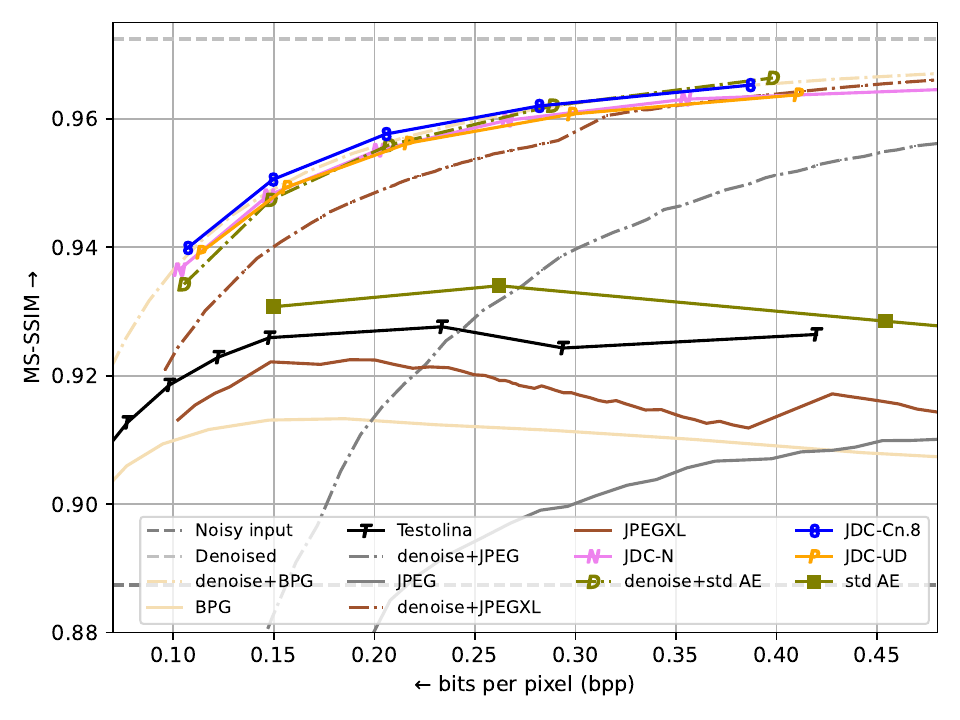}
    \caption{Lossy compression of noisy ($\text{MS-SSIM}\in [0.7, 1.0)$) test images from NIND \cite{nind} with respect to their matching clean ground-truth. This figure combines Figure 3a and Figure 3b in from the text.}
    \label{fig:allplot}
  \end{center}
\end{figure*}


%

%

%

\begin{figure*}
  \begin{center}
    \includegraphics[width=\linewidth]{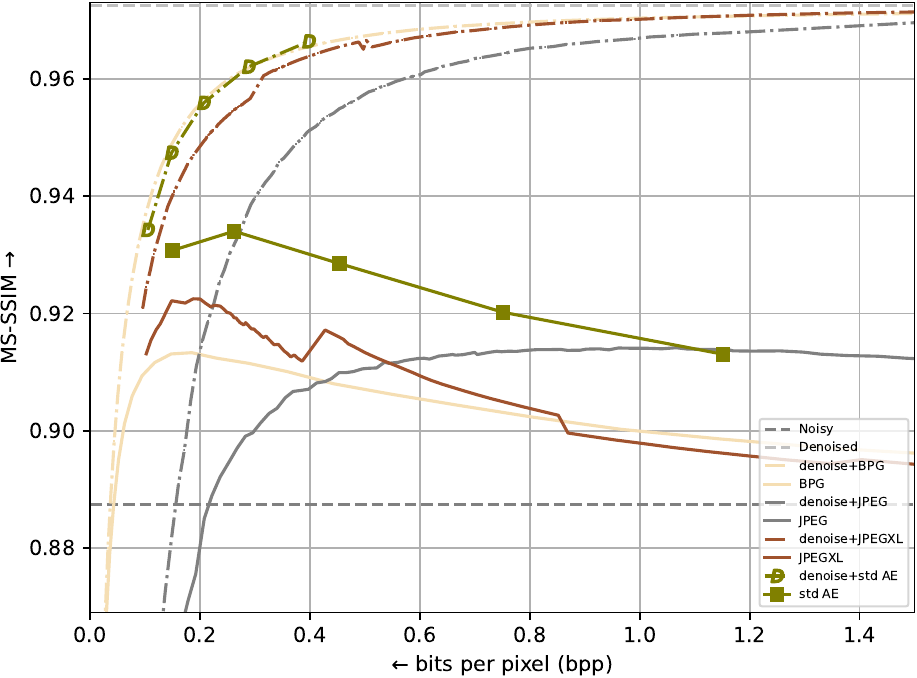}
    \caption{Lossy compression of noisy ($\text{MS-SSIM}\in [0.7, 1.0)$) test images from NIND \cite{nind} with standard methods (JPEG, JPEG XL, BPG, and a standard compression autoencoder \cite{manypriors} abbreviated as ``std AE''): rate-distortion with respect to the clean ground-truth images when encoding noisy images ($\blacksquare$), same compression schemes applied after the test images were denoised with a trained ``universal denoiser'' prior to compression ($D$). Compression alone performs some implicit denoising and the image quality is higher than that of the noisy input given sufficient bitrate, but image quality degrades as the bitrate increases and the noisy signal is eventually reconstructed. Applying a universal denoiser before the compression (dashed lines) greatly improves rate-distortion at the cost of added complexity.}
    \label{fig:denoisethencompress}
  \end{center}
\end{figure*}


%

%


\clearpage
{\small
\bibliographystyle{ieee_fullname}
\bibliography{egbib}
}

\end{document}